\documentclass[letterpaper]{article} 
\usepackage{aaai25}  
\usepackage{times}  
\usepackage{helvet}  
\usepackage{courier}  
\usepackage[hyphens]{url}  
\usepackage{graphicx} 
\usepackage{natbib}  
\usepackage{caption} 
\usepackage{algorithm}
\usepackage{algorithmic}
\usepackage{amsmath}
\usepackage{multirow}
\usepackage{soul, color, xcolor}
\sethlcolor{blue}
\usepackage{newfloat}
\usepackage{listings}
\usepackage{cases}
\DeclareCaptionStyle{ruled}{labelfont=normalfont,labelsep=colon,strut=off} 
\floatstyle{ruled}
\newfloat{listing}{tb}{lst}{}
\floatname{listing}{Listing}
\title{Coherency Improved Explainable Recommendation via Large Language Model}
\author{
    Shijie Liu\textsuperscript{\rm 1}\equalcontrib,
    Ruixing Ding\textsuperscript{\rm 1}\equalcontrib,
    Weihai Lu\textsuperscript{\rm 2}\equalcontrib,
    Jun Wang\textsuperscript{\rm 1},
    Mo Yu\textsuperscript{\rm 3},
    Xiaoming Shi\textsuperscript{\rm 1},
    Wei Zhang\textsuperscript{\rm 1}\thanks{Corresponding author.}
}
\affiliations{
    \textsuperscript{\rm 1}East China Normal University,~~~\\
    \textsuperscript{\rm 2}Peking University,~~~\\
    \textsuperscript{\rm 3}	WeChat AI, Tencent \\
    karrich128@gmail.com, zhangwei.thu2011@gmail.com
}

\usepackage{bibentry}

\begin{document}

\maketitle

\begin{abstract}
Explainable recommender systems are designed to elucidate the explanation behind each recommendation, enabling users to comprehend the underlying logic.
Previous works perform rating prediction and explanation generation in a multi-task manner. 
However, these works suffer from incoherence between predicted ratings and explanations.
To address the issue, we propose a novel framework that employs a large language model (LLM) to generate a rating, transforms it into a rating vector, and finally generates an explanation based on the rating vector and user-item information. 
Moreover, we propose utilizing publicly available LLMs and pre-trained sentiment analysis models to automatically evaluate the coherence without human annotations.
Extensive experimental results on three datasets of explainable recommendation show that the proposed framework is effective, outperforming state-of-the-art baselines with improvements of 7.3\% in explainability and 4.4\% in text quality. 

\end{abstract}
\begin{links}
\link{Code}{https://github.com/karrich/CIER}
\end{links}
\section{Introduction}

Recommendation systems provide personalized suggestions to maximize user engagement and satisfaction based on historical interactions and preferences~\cite{zhang2019deep}, showing significant potential and technological value.
Recently, to relieve the concerns regarding trustworthiness due to the inherent lack of transparency and explainability, 
 explainable recommendation systems have been introduced~\cite{ZhangC20,ZhangYWW22}. 
 These systems elucidate the rationale behind each recommendation, enabling users to comprehend the underlying logic. 
 This enhanced understanding empowers users to make informed decisions and fosters greater trust in the system's suggestions.

\begin{figure}[t]
    \centering
    \includegraphics[width=\linewidth]{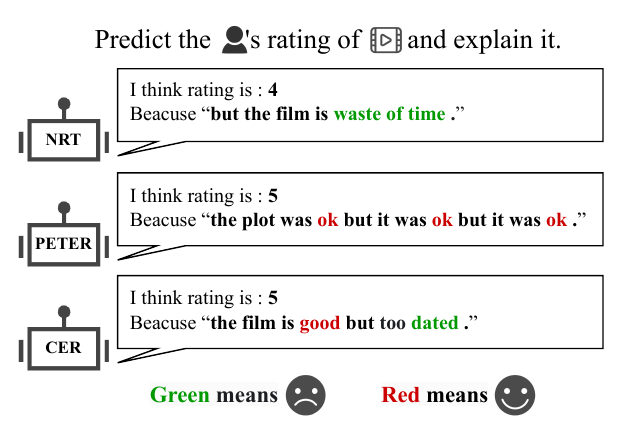}
    \caption{Explanations generated by NRT, PETER, and CER for an example from Amazon Movies.}
    \label{fig:example}
    
\end{figure}

Current works on explainable recommendation systems generate ratings and provide corresponding explanations~\cite{ni-etal-2017-estimating,DualLearning,ACL21-PETER,cheng-etal-2023-explainable}.
Specifically, the rating prediction and explanation generation modules are jointly learned in a multi-task learning manner, 
sharing a common hidden representation layer but having individual output layers.
Despite improvements in explanations, 
these methods suffer from incoherence between predicted ratings and explanations. 
As shown in Figure~\ref{fig:example}, NRT~\cite{NRT} and PETER~\cite{ACL21-PETER} generate inconsistent explanations.
This inconsistency arises because these two tasks only share hidden layer representation, and explanation generation does not explicitly include rating information.
To enhance the coherency, CER~\cite{ECAI23-CER} proposes explanation-based rating estimation, 
obtaining explanation embeddings through max pooling of generated text embeddings and minimizing the distance between the explanation and the corresponding rating vectors.
Despite the improved coherency, 
CER suffers from two issues: (1)
CER utilizes a small-sized transformer as the backbone, which limits the generative performance.
(2) CER struggles to enforce coherence due to poor sentence embedding, as it relies on max pooling of pre-trained word embeddings, which fails to capture rich contextual information~\cite{neelakantan2022text,wang2023improving}.
As such, CER fails to generate coherent explanations, as reflected in the figure.

Recently, the revolutionary progress in \textbf{l}arge \textbf{l}anguage \textbf{m}odels (LLM)~\cite{zeng2022glm,openai2023gpt4,touvron2023llama} has catalyzed substantial technological transformations in natural language generation and reshaped its foundation.
Inspired by LLMs, we propose using them as the backbone model to predict ratings and generate explanations for recommendation systems.
LLMs produce fluent and accurate ratings and explanations, addressing the first issue.
To tackle the second issue, we propose generating ratings and explanations in a pipeline manner, similar to next-token prediction, which is suitable for decoder-based LLMs.

Specifically, an LLM is fine-tuned with LoRA~\cite{lora} to predict ratings, which are subsequently transformed into rating vectors, while explanations are generated using both user and item information in conjunction with rating vectors.
The generation process utilizes the rating as input for the LLM, enhancing the coherency through its in-context learning capability. Meanwhile, training techniques such as rating smoothing, curriculum learning, and multi-task learning are employed to enhance performance, with experiments demonstrating their effectiveness.

Besides, coherency evaluation is crucial yet challenging.
Current methods can be divided into manual and automatic evaluations.
Manual evaluation, while effective, is labor-intensive and impractical at scale. 
To address this,
a study~\cite{ECAI23-CER} proposes using a binary classifier trained on manually annotated data for automatic evaluation.
Despite its high efficiency, this automated metric relies heavily on the quality and quantity of the annotated data, which is time-consuming and costly.
To overcome these limitations, we propose utilizing GPT-4~\cite{achiam2023gpt} and a pre-trained sentiment analysis model~\cite{sentiment} to assess coherency without additional manual annotations. GPT-4 excels in advanced natural language understanding, while the BERT-based pre-trained model is tailored for sentiment classification in product reviews, making both well-suited for our purposes. 

The main contributions are as follows:
\begin{itemize}
    \item To generate more coherent explanations, we propose a framework, named CIER (\textbf{C}oherency-\textbf{I}mproved \textbf{E}xplainable \textbf{R}ecommendation), which initially predicts a rating with LLMs and subsequently leverages the rating to generate an explanation. 

    \item For a more streamlined assessment of coherency between ratings and explanations, we propose to employ LLMs and pre-trained sentiment analysis models.

    \item 
    We conduct extensive experiments to demonstrate the effectiveness of the proposed framework against strong baselines, and experimental results show that training techniques can further improve the results.
    
\end{itemize}

\section{Related Work}

\subsection{Explainable Recommender Systems}
In recent years, more and more research has focused on how to provide good explanations for recommendations to enhance system effectiveness and user satisfaction. Various explanation styles include topical word clouds~\cite{cloud}, highlighted images~\cite{image}, knowledge graphs~\cite{KG}, and automatically generated textual explanations~\cite{ACL21-PETER}. The latter is of particular interest, as textual explanations are more easily comprehended by users, particularly non-expert users, and more informative than pre-defined templates. 

In this work, we focus on generating high-quality explanatory texts while providing accurate recommendations. Our proposed CIER framework aims to address the flaw of inconsistencies between recommendations and natural language explanations provided by existing methods~\cite{ACL21-PETER,NRT,TOIS23-PEPLER,ECAI23-CER,yang2021explanation,zhang2023triple,DualLearning}.

\subsection{LLMs for Explainable Recommendation}
With the advancement of natural language generation techniques, several studies have employed Recurrent Neural Networks (e.g., Long Short-Term Memory~\cite{lstm}, Gated Recurrent Unit~\cite{GRU}), unpretrained Transformer~\cite{transformer} and pre-trained language models (e.g., BERT~\cite{bert}) for generating explanations. Pre-trained large language models are initially introduced in PEPLER~\cite{TOIS23-PEPLER} to enhance the performance of explanation generation. Although PEPLER utilizes prompt-based transfer learning with GPT-2~\cite{Radford2019LanguageMA}, it fails to structure training data in a manner suitable for instruction tuning, thereby limiting the system's ability to produce high-quality explanations. 

Our proposed CIER framework is designed to harness the language capabilities of LLMs to advance the field of explainable recommender systems.

\subsection{Explainable Recommendation Evaluation Metrics}
Previous works mostly rely on perplexity and overlapping-based metrics such as Distinct-N~\cite{distinct-n}, Rouge score~\cite{rouge}, and BLEU score~\cite{bleu}, to evaluate against the ground truth explanations. However, none of these metrics assess how truthfully the generated explanations reflect the rating predictions.

The studies~\cite{ECAI23-CER,yang2021explanation} introduce some automatic methods for evaluating the consistency between predictions and explanations. However, the reliance on the manual rules and quality of annotations significantly impacts the effectiveness and reliability of the evaluation process, which also raises concerns about reproducibility. 
To address these limitations, we introduce a new automatic evaluation method that uses publicly available pre-trained language models to assess rating-explanation coherence.

\begin{figure*}[t]
    \centering
    \includegraphics[width=\linewidth]{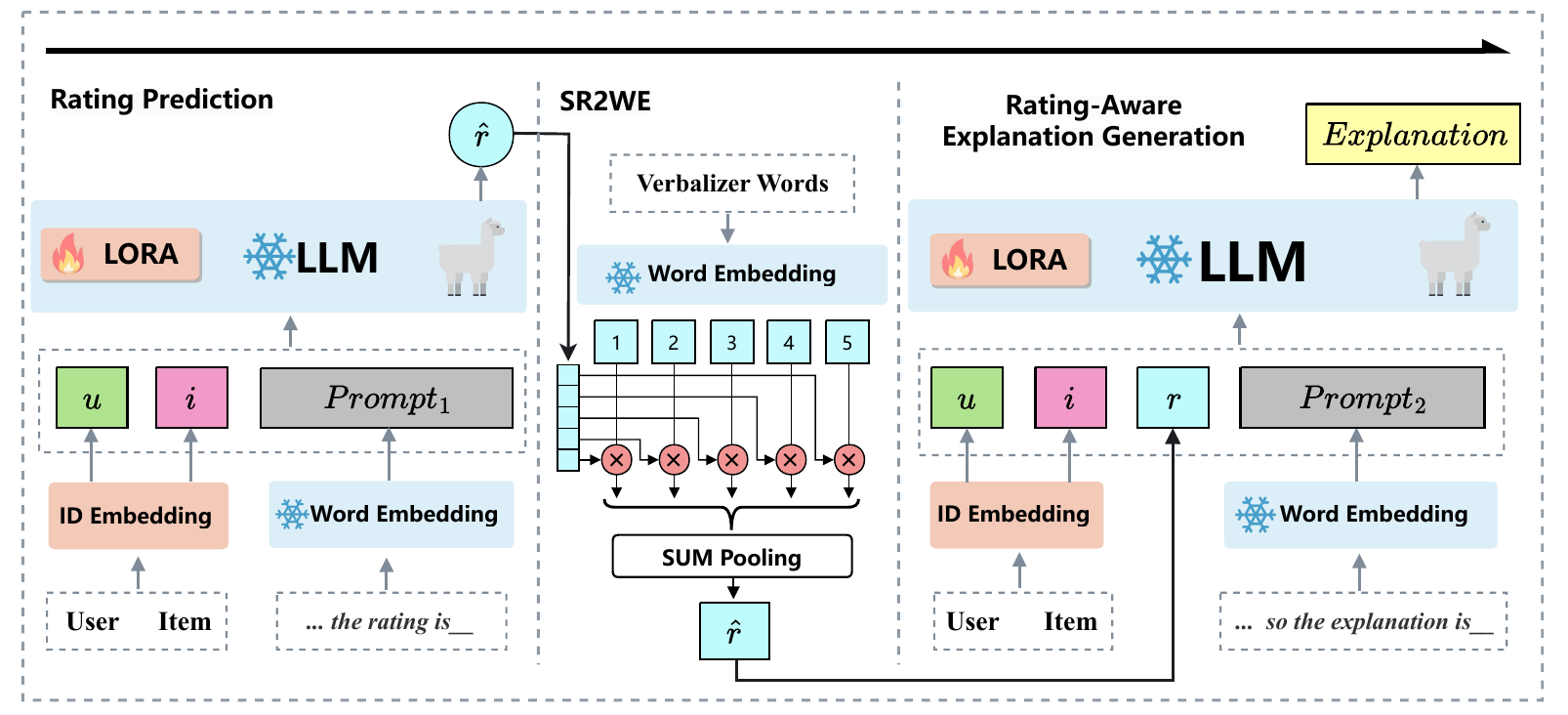}
    \caption{The overview framework of CIER. (a) Rating Prediction: aiming to predict users' ratings of items based on LLM. (b) SR2WE: embedding the predicted soft rating into the LLM word embedding space. (c) Rating-Aware Explanation Generation: using the predicted ratings as context to generate explanations related to the ratings.}
    \label{fig:model}
\end{figure*}

\section{Methodology}
The overview of the proposed method CIER is depicted in Figure~\ref{fig:model}, with three modules, rating prediction, \textbf{s}oft \textbf{r}ating to \textbf{w}ord \textbf{e}mbedding (SR2WE), and explanation generation.
In what follows, we first provide the problem formulation, then introduce the details and training techniques of CIER, and finally describe the proposed automatic evaluation method for assessing the coherence.

\subsection{Problem Formulation}

Given a pair of user $u$ and item $i$, the objective is to jointly predict a rating $r_{u,i}$ and generate an explanation $E_{u,i}$ that justifies this rating. 
The rating $r_{u,i}$ is a score from 1 to 5 that reflects the user $u$'s preference towards the item $i$.
The explanation $E_{u,i}$ is a sequence of tokens from a predefined vocabulary $\mathcal{V}$ that provides a personalized verbalizer.

\subsection{Proposed Method CIER}
\subsubsection{Rating Prediction}
The objective of the rating prediction task is to estimate the rating a user $u$ would give to an item $i$, similar to typical recommendation tasks. To construct a unified framework for both the rating prediction and explanation generation tasks, we employ LLaMA2-7B as the backbone for CIER and use a corresponding verbalizer~\cite{hu2021knowledgeable} specifically for the rating prediction component.

The verbalizer $V^r$ is a fixed mapping from numeric ratings to their word representations, defined as $V^r=$\{1: “1”, 2: “2”, 3: “3”, 4: “4”, 5: “5”\}. This design facilitates consistency in the rating prediction process.
The probability assigned by the model to each word in $V$ corresponds to the probability of each respective rating:
\begin{equation}
\begin{aligned}
\hat{r}_{u,i} = f([u, i, p_{1}, \ldots, p_{m}]),
\end{aligned}
\end{equation}
where $p$ represents the prompt, $f$ is the LLM, $m$ is the prompt length, and $\hat{r}_{u,i}$ is the predicted probability of each rating. 
Then the rating is obtained by weighted summation:
\begin{equation}
\begin{aligned}
\hat{r}_{score} = \sum_{x=1}^{|r|} \hat r_{u,i,x} \cdot x ~,
\end{aligned}
\end{equation}
where $|r|$ is the number of rating classes, $\hat r_{u,i,x}$ is the probability of rating $x$, and \( \sum_{x=1}^{|r|} \hat r_{u,i,x} = 1 \).

\subsubsection{Soft Rating to Word Embedding}

For a given rating, the hard rating embedding directly uses the corresponding word embedding in the verbalizer.
However, hard ratings have less information than soft ratings, so we try to embed soft-rating into the word embedding space, which is defined as follows:
\begin{equation}
\begin{aligned}
\mathbf{s}_{r_{u,i}} = \sum_{x=1}^{|r|} \hat r_{u,i,x} \cdot \text{Embedding}_{LLM}(V^r(x)) ~,
\end{aligned}
\end{equation}
where $\text{Embedding}_{LLM}$ is the word embedding layer of the LLM, and $V^r(x)$ is the corresponding word of rating $x$ in the verbalizer.
At this point, we have obtained the semantic representation of the predicted rating, which encapsulates the uncertainty and distribution features of user $u$'s preference towards item $i$.

\subsubsection{Rating-Aware Explanation Generation}
The rating-aware explanation generation module aims to generate an explanation based on given $u$, $i$, and $r_{u,i}$.

The process is formulated as follows:
\begin{equation}
\begin{aligned}
E_{u,i} = f([u, i, {s}_{r_{u,i}}, p_{1}, \ldots, p_{j}]),
\end{aligned}
\end{equation}
where $p$ represents the prompt, $f$ is the LLM, $j$ is the prompt length, ${s}_{r_{u,i}}$ is the rating embedding from SR2WE module, and $E_{u,i}$ is the generated explanation.

\subsection{Training Techniques}
To balance efficiency and performance, we conduct Lora tuning for LLM. In addition, three training techniques are utilized in this work for better performance, i.e., rating smoothing, curriculum learning, and multi-task learning.
\subsubsection{Rating Smoothing}
Using a probability distribution over possible ratings to obtain the rating embedding in the inference phase offers several potential benefits.
However, training the model exclusively on ground-truth ratings introduces a notable disparity between the training phase and inference. 

To address this, we introduce a rating smoothing technique that is inspired by label smoothing but incorporates enhancements tailored to our specific scenario. Traditional label smoothing distributes probability across all categories, potentially diluting the model's sensitivity to user-specific ratings.
In rating prediction, adjacent ratings contain similar sentiments, so our proposed rating smoothing prevents over-smoothing by limiting the impact to ratings that are numerically adjacent to the ground truth ratings (called neighboring ratings).
Specifically, 
with a probability \(\gamma\), the original one-hot distribution of rating $r_{u,i}$ is transformed to:
\begin{equation}
\begin{aligned}
{r}_{u,i,x}^{\text{modified}} = 
\begin{cases} 
{1 - \alpha} & \text{if } x = r_{u,i} \\
\frac{\alpha}{k} & \text{if } x \in \mathcal{N}_{r_{u,i}}^k\\
0 & \text{others} \,,
\end{cases} 
\end{aligned}
\end{equation}
where \(\alpha \in [0, \frac{k}{k+1}]\), \(\mathcal{N}_{r_{u,i}}^k\) denotes the set of \(k\) neighboring ratings of $r_{u,i}$.
Regarding to the selection of the smoothing technique, various possibilities are explored and the proposed rating smoothing is intuitive and experimentally proven to be effective.

\subsubsection{Training With Curriculum Learning}

Previous methods struggle to capture explanatory keywords that reflect users' interests in explanations, showing low explainability. 
To address this problem, we introduce a keyword generation task to help the model identify item features (e.g. lobby, location) that the user cares about in explanations. 

Inspired by curriculum learning, we propose a training strategy that allows models to build foundational knowledge before tackling more intricate problems. 
Specifically, we devise a linear transition mechanism that dynamically adjusts the data allocation between the keyword generation and explanation generation tasks during training. 
The transition probability $P(t)$ represents the likelihood of the data point used for explanation generation task in batch $t$:
\begin{equation}
\begin{aligned}
P(t) = \frac{t}{T} ~,
\end{aligned}
\end{equation}
where $T$ denotes the total number of training batches.
During each batch, data points are probabilistically assigned to either task based on a random number 
$n$ generated from a uniform distribution over [0, 1].
The assignment is determined by comparing \( n \) with \( P(t) \):

\begin{subnumcases} {\label{weqn} \text{Task}(t) =}
Task_{\text{explanation}} & \text{if } $n < P(t)$ \\
Task_{\text{keyword}} & \text{if } $n \geq P(t)$\,.
\end{subnumcases}
The training process initially focuses on predicting the keywords of explanations, gradually shifting towards generating complete explanations.
This approach retains foundational knowledge while integrating the complexities of explanation generation. 
\begin{figure}[t!]
    \centering
    \includegraphics[width=\linewidth]{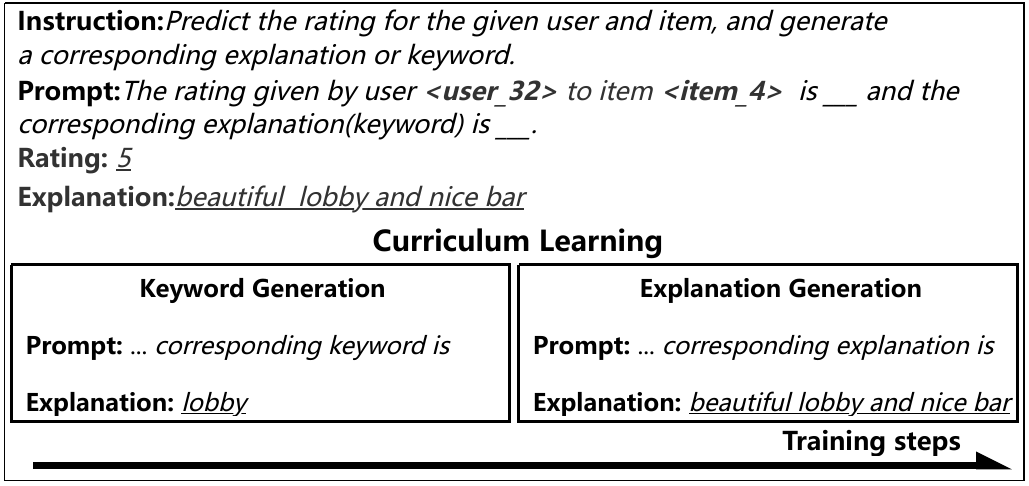}
    \vspace{-1em}
    \caption{Instructions and prompts for curriculum learning.}
    \label{fig:curriculum learning}
\end{figure}

Figure~\ref{fig:curriculum learning} shows the specific instructions and prompts used. During the keyword training process, ``explanation'' in the prompt will be replaced with ``keyword'', and the target will be replaced from a complete explanation to the key words in the explanation.

\subsubsection{Multi-Task Learning}

The cross-entropy loss (CE) is utilized as the loss function for rating prediction:
\begin{equation}
\begin{aligned}
\mathcal{L}_r = - \frac{1}{|\mathcal{T}|} \sum_{(u,i) \in \mathcal{T}} \sum_{x=1}^{|r|} r_{u,i,x} \log(\hat{r}_{u,i,x}) ~,
\end{aligned}
\end{equation}
where $\mathcal{T}$ denotes the training set and $r_{u,i,x}$ is the probability of the ground-truth rating being \( x \).
We use the Negative Log-Likelihood (NLL) as the loss function for the text (i.e., explanation or keyword) generation, computing the mean over user-item pairs in the training set.
\begin{equation}
    \begin{aligned}
        \mathcal{L}_e = \frac{1}{|\mathcal{T}|} \sum_{(u,i) \in \mathcal{T}} \frac{1}{|E_{u,i}|} \sum_{t=1}^{|E_{u,i}|} -\log c_{|S| - |E_{u,i}| + t}^{e_t} ~.
    \end{aligned}
\end{equation}
The probability $c_t^{e_t}$ is offset by $|S| - |E_{u,i}| + t$ positions because the
generated text is placed at the end of the sequence. 

We integrate rating prediction and text generation into a multi-task learning framework. The objective function is defined as follows: 
\begin{equation}
\begin{aligned}
\mathcal{J} = \min_{\Theta = \{\Theta_{Lora}, \Theta_{U}, \Theta_{I}\}} (\mathcal{L}_e + \lambda \mathcal{L}_r) ~,
\end{aligned}
\end{equation}
where $\Theta$ denotes all the trainable parameters in the model, including the parameters of Lora modules, i.e., $\Theta_{Lora}$, and the parameters of ID Embeddings, i.e., $\Theta_{U}$ and $\Theta_{I}$. The hyperparameter $\lambda$ is used to balance the learning between the explanation generation task and the rating prediction task. 
It is worth noting that these two tasks are performed in a pipeline manner like next-token prediction, thus suitable for decoder-based LLMs.

\subsection{Automatic Coherence Evaluation}
\label{automatic coherence}
To address heavy reliance on high-quality annotated data in the previous approach~\cite{ECAI23-CER}, we employ publicly available pre-trained language models, specifically GPT-4 and bert-base-multilingual-uncased-sentiment~\cite{sentiment}, to automatically assess the rating-explanation coherency. 
GPT-4 has recently demonstrated remarkable performance across various tasks, leading to its widespread use as an evaluator~\cite{10.5555/3666122.3666237,NEURIPS2023_ac662d74}. Meanwhile, the BERT-based model is specifically designed for sentiment analysis in product reviews, making it particularly suitable for our purposes.

For GPT-4, a prompt is utilized to provide clear guidelines on how sentiment should match each rating level and
an instruction is used to make it respond with ``Yes'' or ``No'' based on the coherency between ratings and explanations. 
The percentage of coherent rating-explanation pairs identified by GPT-4 serves as a performance metric. 
Specifically, the ``gpt-4o'' model is utilized to evaluate randomly sampled 500 predictions from each model. 

Bert-base-multilingual-uncased-sentiment is applied to predict the sentiment rating of explanations for all predictions. 
Given the influence of personalized factors on rating predictions and the individual biases across different datasets, coherency is defined as the predicted sentiment rating deviating by no more than one point from the given rating, defined as follows:
\begin{subnumcases}
{\text{Coherency} =}
1 & \text{if}  $| y - \hat{y}| \leq 1$ \\
0 & \text{otherwise}
\end{subnumcases}
where \( y \) represents the rating provided by explainable recommendation model, and  \( \hat{y} \) represents that from the sentiment classification model.

\section{Experiments}
\subsection{Experimental Setting}
\subsubsection{Dataset}
To validate the effectiveness of our method, we conducted experiments on three publicly available datasets and their splits~\cite{nete}. Each dataset is randomly divided into training, validation, and test sets in an 8:1:1 ratio five times. The three datasets are from TripAdvisor (hotel), Amazon (movies \& TV), and Yelp (restaurant). Each record in the dataset consists of a user ID, an item ID, a rating on a scale of 1 to 5, an explanation, and item features. The explanations are sentences extracted from user reviews. 
Features are attributes of items extracted from the explanation, e.g., \textit{lobby}, which represent aspects users care about, and we consider them as the keyword of explanations. 
The dataset statistics are shown in Table~\ref{tab:statistics}.
The available datasets and keyword extraction tools are provided by Sentires~\cite{SIGIR14-Sentires, Sentires}.

\begin{table}[t]
    \small
    \centering
    \begin{tabular}{l|r|r|r}
        \hline
        Datasets&Yelp&Amazon&TripAdvisor\\
        \hline
        \#users&27,147&7,506&9,765\\
        \#items&20,266&7,360&6,280\\
        \#records&1,293,247&441,783&320,023\\
        \#features&7,340&5,399&5,069\\
        \hline
    \end{tabular}
    \vspace{-0.5em}
    \caption{Statistics of the datasets.}
    \label{tab:statistics}
\end{table}
\subsubsection{Evaluation Metrics}
To evaluate the performance of rating prediction, we utilize two commonly used metrics: Root Mean Square Error (\textbf{RMSE}) and Mean Absolute Error (\textbf{MAE}) to measure the deviation between predicted ratings and ground truth ratings.

For explanation performance, we measure the generated explanations from two main perspectives: text quality and explainability. For the text quality, we use \textbf{BLEU}~\cite{bleu} and \textbf{ROUGE}~\cite{rouge}, which are common metrics in natural language generation tasks.
Specifically, we use BLEU-1 and BLEU-4 metrics to evaluate the precision, the recall-scores of ROUGE-1 and ROUGE-2 to evaluate the recall, and the f1-score of ROUGE-L for comprehensive evaluation.
For the text explainability, we use additional indicators proposed by ~\cite{nete} to measure explainability: Feature Matching Ratio (\textbf{FMR}), Feature Coverage Ratio (\textbf{FCR}), Feature Diversity (\textbf{DIV}), and Unique Sentence Ratio (\textbf{USR}).

To measure the coherence between explanations and predicted ratings, we perform manual and automated evaluations. For manual evaluation, we follow CER~\cite{ECAI23-CER} to annotate the coherence with two independent human annotators. For automatic annotation, we use our proposed automatic evaluation method.

\subsubsection{Baselines}
\begin{table*}[ht!]
\small
    \begin{center}
            \begin{tabular}{l|cccc|ccccc}
                \hline
                \multirow{2}{*}&\multicolumn{4}{c|}{\textbf{Explainability}}&\multicolumn{5}{c}{\textbf{Text Quality}}\\
                \cline{2-10}
                &FMR$\uparrow$&FCR$\uparrow$&DIV$\downarrow$&USR$\uparrow$&B-1$\uparrow$&B-4$\uparrow$&R-1$\uparrow$&R-2$\uparrow$&R-L$\uparrow$\\
                \hline
                \hline
                &\multicolumn{9}{c}{Yelp}\\
                \hline
                NRT~\cite{NRT}&6.65&11.96&1.77&16.02&11.36&0.65&12.35&1.29&10.39\\
                Att2Seq~\cite{ATT2Seq}&7.08&14.24&1.72&17.65&11.47&0.69&12.46&1.35&10.40\\
                PETER~\cite{ACL21-PETER}&8.09&13.80&\underline{1.65}&8.47&9.68&0.62&11.63&1.26&10.24\\
                CER~\cite{ECAI23-CER}&8.05&15.00&\textbf{1.59}&9.67&10.03&0.65&11.72&1.29&10.27\\
                PEPLER~\cite{TOIS23-PEPLER}&8.11&21.04&1.73&20.32&10.94&0.67&12.05&1.36&10.41\\
                ERRA$^{\ast}$~\cite{cheng-etal-2023-explainable}&\textbackslash&\textbackslash&\textbackslash&\textbackslash&10.71&0.73&\textbackslash&1.36&10.82\\
                \hline
                CIER (Curriculum Learning)&\textbf{8.71}&\textbf{53.84}&1.67&\textbf{32.63}&\textbf{11.78}&\textbf{0.83}&\textbf{13.02}&\textbf{1.59}&\textbf{10.90}\\
                ~~~~~~~~~~~~~Two-Stage Learning&8.61&\underline{52.46}&1.67&{31.30}&\underline{11.62}&\underline{0.82}&\underline{12.89}&\underline{1.57}&\underline{10.85}\\

                ~~~~~~~~~~~~~Vanilla Learning&\underline{8.62}&52.19&1.70&\underline{32.48}&11.39&0.79&12.78&1.54&\underline{10.85}\\
                \hline
                \hline
                
                &\multicolumn{9}{c}{Amazon}\\
                \hline
                NRT~\cite{NRT}&11.13&5.67&2.38&14.57&12.62&0.89&13.82&1.88&11.24\\
                Att2Seq~\cite{ATT2Seq}&11.11&8.22&2.17&22.12&12.86&0.92&13.88&1.87&11.19\\
                PETER~\cite{ACL21-PETER}&11.60&9.12&2.20&13.30&12.38&1.00&13.45&1.94&11.29\\
                CER~\cite{ECAI23-CER}&11.47&10.25&2.09&14.72&12.02&1.02&13.23&1.92&11.05\\
                PEPLER~\cite{TOIS23-PEPLER}&11.88&34.07&2.26&24.87&12.57&1.03&13.83&1.92&11.31\\
                \hline
                CIER (Curriculum Learning)&\textbf{12.45}&\textbf{51.80}&\underline{2.08}&46.99&\textbf{13.55}&{1.15}&\textbf{14.61}&{2.09}&\textbf{11.70}\\

                ~~~~~~~~~~~~~Two-Stage Learning&\underline{12.21}&\underline{51.11}&\textbf{2.01}&\underline{50.26}&13.43&\underline{1.18}&14.50&\underline{2.12}&\underline{11.67}\\
                
                ~~~~~~~~~~~~~Vanilla Learning&{12.00}&50.84&\underline{2.08}&\textbf{50.92}&\underline{13.53}&\textbf{1.23}&\underline{14.58}&\textbf{2.13}&11.65\\
                \hline
                \hline
                &\multicolumn{9}{c}{TripAdvisor}\\
                \hline
                NRT~\cite{NRT}&5.76&14.15&3.09&18.29&14.85&0.96&15.07&1.98&12.24\\
                Att2Seq~\cite{ATT2Seq}&5.78&10.61&\underline{2.92}&10.33&15.16&0.97&15.17&1.97&12.22\\
                PETER~\cite{TOIS23-PEPLER}&6.47&13.72&{3.03}&9.60&15.97&1.04&15.94&2.25&12.64\\
                CER~\cite{ECAI23-CER}&6.97&12.99&3.14&9.18&15.59&1.09&15.89&2.19&12.75\\
                PEPLER~\cite{TOIS23-PEPLER}&7.36&19.91&3.35&{24.29}&15.06&1.02&14.92&2.03&12.21\\
                ERRA$^{\ast}$~\cite{cheng-etal-2023-explainable}&\textbackslash&\textbackslash&\textbackslash&\textbackslash&16.13&1.06&\textbackslash&2.15&13.17\\
                \hline
                CIER (Curriculum Learning)&\textbf{8.08}&\underline{36.99}&3.05&\underline{29.86}&\textbf{17.00}&\textbf{1.31}&\textbf{17.07}&\textbf{2.54}&\textbf{13.40}\\

                ~~~~~~~~~~~~~Two-Stage Learning&\underline{7.89}&\textbf{39.08}&3.00&\textbf{31.80}&\underline{16.54}&\underline{1.28}&\underline{16.70}&\underline{2.45}&\underline{13.33}\\
                
                ~~~~~~~~~~~~~Vanilla Learning&{7.73}&{36.60}&\textbf{2.86}&{27.63}&{16.45}&{1.25}&{16.66}&{2.40}&{13.31}\\
                \hline
                \hline

            \end{tabular}
    \end{center}
    \vspace{-1em}
    \caption{Results of explanation. B-1, B-4, R-1, R-2 and R-L represent the scores of BLUE-1, BLEU-4, ROUGE-1, ROUGE-2 and ROUGE-L, respectively. BLEU, ROUGE, FMR, FCR, and USR are presented as percentage (\%), while the others are absolute values. The best values in the table are represented in bold, and the second-best values are represented with underlines. Stars$^{\ast}$ indicate that the results of this method are from its paper.}
    \label{tab:explanation}
\end{table*}

\begin{table}[t]
\small
    \begin{center}
        \resizebox{\linewidth}{!}{
            \begin{tabular}{llc}
                \hline
                Method&Explanation&rating\\
                \hline
                Truth&swimming \textbf{pool} was small and shallow&1\\
                NRT&the bed was comfortable and the room was comfortable&3\\
                PETER&the hotel is a little dated but the rooms are very small&3\\
                CER&the resort is a bit dated but the hotel is a bit dated&1\\
                \textbf{CIER}&the \textbf{pool} is a bit small and the gym is a bit small&1\\
                \hline
            \end{tabular}
        }
    \end{center}
    \vspace{-1em}
    \caption{Example generated by CIER and baselines.}
    \label{tab:output}
\end{table}

To evaluate the explainability performance, we compare the following explanation methods:

\noindent\textbf{NRT}~\cite{NRT} utilizes GRU~\cite{GRU} to jointly predict ratings and generate explanations using user and item IDs as input. 

\noindent\textbf{Att2Seq} ~\cite{ATT2Seq} is an explanation generation model based on LSTM~\cite{lstm}. 

\noindent \textbf{PETER} ~\cite{ACL21-PETER} is a powerful multi-layer Transformer~\cite{transformer} model that simultaneously predicts ratings and generates explanations.

\noindent\textbf{CER} ~\cite{ECAI23-CER} proposes a module that estimates the discrepancy between predicted ratings and explanation-based ratings to enhance rating-explanation coherency. 

\noindent\textbf{PEPLER} ~\cite{TOIS23-PEPLER} leverages the advanced capabilities of GPT-2 through prompt-based transfer learning and regularization loss. 

\noindent\textbf{ERRA} ~\cite{cheng-etal-2023-explainable} is a multi-layer Transformer with aspect enhancement and retrieval enhancement. Since the code is incomplete, we directly use its results in its paper.

For the evaluation of recommendation performance, in addition to NRT, PETER, and CER, we also use three traditional models as baselines:

 
\noindent\textbf{SVD++} ~\cite{SVD++} integrates implicit feedback from users to enhance the latent factors.

\noindent\textbf{DeepCoNN} ~\cite{DeepCoNN} learns item properties and user behavior from review text.

\noindent\textbf{NARRE} ~\cite{NARRE}  applies the attention mechanism to the rating prediction task.

For evaluating coherence, we use PETER, CER, and CIER-M as baselines. CIER-M means that CIER masks the context (predicted ratings) when generating explanations.

\subsection{Implementation Details}
All the experiments are conducted on an NVIDIA H800 GPU.
We utilize the validation set to tune hyper-parameters for each dataset, and subsequently present the average evaluation metrics computed across 5 data splits on the testing set. 
We load LLaMA2-7B from HuggingFace as the backbone of our proposed model, utilizing BPE~\cite{BPE} for vocabulary construction. 
To ensure fair comparisons, we apply BPE to all baseline models and set the max explanation length to 20 BPE tokens.
For CIER, $\lambda$ is set to 0.1 and $\gamma$ to 0.2, selected through grid search over the ranges \([0.01, 0.1, 1.0, 10.0]\) and \([0.0, 0.2, 0.5, 0.8, 1.0]\), respectively.
The model is optimized using the AdamW ~\cite{Adamw} optimizer with hierarchical learning rates: $10^{-4}$ for the Lora module and $10^{-3}$ for the other components. The training epoch is set to 3 and the embedding size $d$ is set to 1024. 
At the end of each epoch, we calculate the model's loss on the validation set. If the validation loss does not decrease anymore, the model is saved.

\begin{table}[!t]
\normalsize
    \begin{center}
        \resizebox{\linewidth}{!}{
            \begin{tabular}{r|cc|cc|cc}
                \hline
                \multirow{2}{*}&\multicolumn{2}{c|}{Yelp}&\multicolumn{2}{c|}{Amazon}&\multicolumn{2}{c}{TripAdvisor}\\
                \cline{2-7}
                &R$\downarrow$&M$\downarrow$&R$\downarrow$&M$\downarrow$&R$\downarrow$&M$\downarrow$\\
                \hline
                SVD++&1.019&0.791&0.965&0.722&0.809&0.617\\
                
                DeepCoNN& 1.108& 0.883& 1.108& 0.881&0.888&0.683\\
                NARRE& 1.031& 0.811& 1.003& 0.780&0.817&0.622\\
                NRT&1.016&0.796&0.954&\underline{0.706}&\textbf{0.791}&\textbf{0.605}\\
                PETER&\underline{1.013}&\underline{0.783}&0.953&0.709&0.806&0.623\\
                CER&\underline{1.013}&0.787&\underline{0.952}&0.713&0.814&0.637\\
                CIER&\textbf{1.009}&\textbf{0.781}&\textbf{0.951}&\textbf{0.705}&\underline{ 0.797}&\underline{ 0.612}\\
                \hline
            \end{tabular}
        }
    \end{center}
    \vspace{-1em}
    \caption{The comparison of the recommendation performance of CIER and other baseline methods. “R” means RMSE and “M” means MAE.}
    \label{tab:recommendation}
\end{table}

\begin{table*}[!t]
\small
    \begin{center}
            \begin{tabular}{r|ccc|ccc|ccc}
                \hline
                 \multirow{2}{*}&\multicolumn{3}{c|}{GPT-4}&\multicolumn{3}{c|}{Sentiment-Bert}&\multicolumn{3}{c}{Human annotators}\\
                \cline{2-10}
                &Yelp&Amazon&TripAdvisor&Yelp&Amazon&TripAdvisor&Yelp&Amazon&TripAdvisor\\
                \hline
                PETER&87.2&79.6&87.0&69.2&69.6&80.3&62.0&63.0&84\\
                CER&88.8&80.0&90.4&70.1&70.6&80.7&65.6&66.0&82.5\\
                CIER-M&87.0&81.8&88.0&69.8&74.1&80.1&69.5&60.5&83.5\\
                CIER &\textbf{90.2}&\textbf{89.8}&\textbf{91.6}&\textbf{70.6}&\textbf{77.6}&\textbf{82.2}&\textbf{73.5}&\textbf{74.0}&\textbf{87.0}\\
                \hline
            \end{tabular}
    \end{center}
    \vspace{-1em}
    \caption{Results of coherence evaluation using GPT-4, BERT-based sentiment classification models and human annotations for explanations and prediction ratings of the selected methods.}
    \label{tab:auto}
\end{table*}
\subsection{Evaluation of Explanation}

The text quality and explainability of various explanation generation methods are presented in Table~\ref{tab:explanation}. 
In terms of text quality, our proposed CIER consistently outperforms the baselines on different datasets, demonstrating its effectiveness in generating high-quality sentences.
Table~\ref{tab:output} presents an example generated by the CIER model and some baselines.
By referring to the ground-truth explanation, CIER produces a more accurate explanation.

Regarding explainability, CIER consistently outperforms the baselines on FMR, FCR, and USR, indicating it effectively captures key information in explanatory texts. PEPLER and CIER with vanilla learning, while not explicitly optimized for explainability, demonstrate competitive performance. This can be attributed to their inherent text generation ability obtained by pre-training, enabling them to focus on the nuances and key information within explanations.

\subsection{Evaluation of Rating Prediction}
Evaluation of recommendation accuracy is shown in Table~\ref{tab:recommendation}. 
The experimental results indicate that the proposed method, leveraging an LLM backbone, exhibits strong recommendation performance across all datasets, especially excelling in larger datasets (i.e., Yelp and Amazon). In the smaller, sparser TripAdvisor dataset, while traditional models like NRT perform better, our method still outperforms other Transformer-based models (i.e., PETER and CER).

\subsection{Evaluation of Coherence}

The evaluation of the coherence between the explained and predicted ratings is shown in Table~\ref{tab:auto}. The manual annotation was performed by two volunteers who selected 100 data points from each dataset for the selected methods.
Before annotation, the agreement between the two instructed annotators was measured using the kappa coefficient on a random sample of 200 data points, resulting in a score of 0.918.

Our approach consistently maintains significant advantages in coherence. In particular, our method consistently outperforms CIER-M, suggesting that our approach of using predicted ratings to guide explanation generation allows the model to understand the relationship between ratings and explanations, thereby improving the relationship between explanations and predicted ratings.

\subsection{Effect of Keyword Generation Task}
To test the effect of our designed keyword generation task, we experimented with three different learning strategies:
\begin{itemize}
    \item[1)] \textbf{Vanilla Training},
    which involves training solely for rating prediction and explanation generation.
    While straightforward, it struggles to capture key explanatory words in explanations.
    \item[2)] \textbf{Two-Stage Training}, which involves the model first learning to generate keywords before shifting to explanation generation. 
    While this process helps build a solid foundation, it risks the model forgetting keyword generation knowledge.
    \item[3)] \textbf{Curriculum Learning (Ours)}, which employs a gradual transition from keyword generation to explanation generation.
    It reduces the risk of forgetting keyword generation knowledge and minimizes its negative impacts.
\end{itemize}

\begin{table}[t]
\small
    \begin{center}
        \resizebox{\linewidth}{!}{
            \begin{tabular}{l|cc|cc|cc}
                \hline
                \multirow{2}{*}&\multicolumn{2}{c|}{Yelp}&\multicolumn{2}{c|}{Amazon}&\multicolumn{2}{c}{TripAdvisor}\\
                \cline{2-7}
                &FMR$\uparrow$&R-L$\uparrow$&FMR$\uparrow$&R-L$\uparrow$&FMR$\uparrow$&R-L$\uparrow$\\
                \hline
                CIER&\textbf{8.71}&\textbf{10.90}&\textbf{12.45}&\textbf{11.70}&\textbf{8.08}&\textbf{13.40}\\
                \hline

                w/o RS&8.66&10.80&12.35&11.65&7.95&13.31\\
                
                w/o SR2WE&\underline{8.67}&\underline{10.86}&\underline{12.40}&\underline{11.68}&\underline{8.03}&\underline{13.37}\\
                w/o RA&8.58&10.72&12.36&11.59&7.92&13.35\\
                \hline
            \end{tabular}
        }
    \end{center}
    \vspace{-1em}
    \caption{Ablation analysis of explanation tasks. ``RS'' means rating smoothing, ``RA'' means rating-aware.}
    \label{tab:ablation}
\end{table}

All training processes consist of 3 epochs. For Two-Stage Training, the epochs are distributed in a ratio of 1:2 between the first and second stages.  
The experimental results are shown in Table~\ref{tab:explanation}. 
The two-stage training strategy fails to improve the explainability on the Yelp dataset, possibly due to its large size, which led to knowledge forgetting. Curriculum Learning strategy demonstrates the best performance across all datasets.
It makes the model effectively retain and utilize learned knowledge on keyword generation, resulting in more relevant and accurate explanations. 
However, curriculum learning does not achieve the best performance on the Amazon dataset, likely because 50\% of its keywords appear only once, 10\% more than that in the other datasets.
Thus it is harder to use keywords for generation.

\subsection{Ablation Study}

Table~\ref{tab:ablation} provides the results of the ablation experiments. After removing Rating Smoothing, both the explainability and text quality decline across all datasets.  

Moreover, removing the SR2WE and inserting the ratings from rating smoothing directly into the LLM prompts through the linear layer leads to a decrease in model performance, which indicates that the SR2WE module could better embed the ratings into the word vector space.

After disabling Rating-Aware generation, all indicators show a significant decline, indicating that explicit use of rating information is very beneficial for explanation generation.

\section{Conclusion}
In this paper, we introduce a novel method that utilizes LLMs as the backbone generation model, predicting ratings and explanations with some tailored training techniques. Additionally, we propose to employ LLMs and pre-trained sentiment analysis models to automatically evaluate the coherency between ratings and explanations. Extensive experimental results demonstrate that our approach outperforms the previous state-of-the-art approaches.

\section{Acknowledgments}
This work was supported in part by National Natural Science Foundation of China ( No. 62072182 and No. 92270119), Shanghai Institute of Artificial Intelligence for Education, and Key Laboratory of Advanced Theory and Application in Statistics and Data Science, Ministry of Education.

\bibliography{aaai25}

\begin{thebibliography}{43}
\providecommand{\natexlab}[1]{#1}

\bibitem[{Achiam et~al.(2023)Achiam, Adler, Agarwal, Ahmad, Akkaya, Aleman, Almeida, Altenschmidt, Altman, Anadkat et~al.}]{achiam2023gpt}
Achiam, J.; Adler, S.; Agarwal, S.; Ahmad, L.; Akkaya, I.; Aleman, F.~L.; Almeida, D.; Altenschmidt, J.; Altman, S.; Anadkat, S.; et~al. 2023.
\newblock Gpt-4 technical report.
\newblock \emph{arXiv preprint arXiv:2303.08774}.

\bibitem[{Al-Taie and Kadry(2014)}]{cloud}
Al-Taie, M.; and Kadry, S.~N. 2014.
\newblock Visualization of Explanations in Recommender Systems.
\newblock \emph{Journal of Advanced Management Science}, 2: 140--144.

\bibitem[{Chen et~al.(2018)Chen, Zhang, Liu, and Ma}]{NARRE}
Chen, C.; Zhang, M.; Liu, Y.; and Ma, S. 2018.
\newblock Neural attentional rating regression with review-level explanations.
\newblock In \emph{Proceedings of the 2018 world wide web conference}, 1583--1592.

\bibitem[{Chen et~al.(2019)Chen, Chen, Xu, Zhang, Cao, Qin, and Zha}]{image}
Chen, X.; Chen, H.; Xu, H.; Zhang, Y.; Cao, Y.; Qin, Z.; and Zha, H. 2019.
\newblock Personalized Fashion Recommendation with Visual Explanations based on Multimodal Attention Network: Towards Visually Explainable Recommendation.
\newblock In \emph{SIGIR}, 765–774.

\bibitem[{Cheng et~al.(2023)Cheng, Wang, Lu, Zhang, Zhou, Lu, and Liao}]{cheng-etal-2023-explainable}
Cheng, H.; Wang, S.; Lu, W.; Zhang, W.; Zhou, M.; Lu, K.; and Liao, H. 2023.
\newblock Explainable Recommendation with Personalized Review Retrieval and Aspect Learning.
\newblock In Rogers, A.; Boyd-Graber, J.; and Okazaki, N., eds., \emph{ACL}, 51--64.

\bibitem[{Cho et~al.(2014)Cho, van Merri{\"e}nboer, Gulcehre, Bahdanau, Bougares, Schwenk, and Bengio}]{GRU}
Cho, K.; van Merri{\"e}nboer, B.; Gulcehre, C.; Bahdanau, D.; Bougares, F.; Schwenk, H.; and Bengio, Y. 2014.
\newblock Learning Phrase Representations using {RNN} Encoder{--}Decoder for Statistical Machine Translation.
\newblock In \emph{EMNLP}, 1724--1734.

\bibitem[{Devlin et~al.(2019)Devlin, Chang, Lee, and Toutanova}]{bert}
Devlin, J.; Chang, M.-W.; Lee, K.; and Toutanova, K. 2019.
\newblock {BERT}: Pre-training of Deep Bidirectional Transformers for Language Understanding.
\newblock In Burstein, J.; Doran, C.; and Solorio, T., eds., \emph{NAACL}, 4171--4186.

\bibitem[{Dong et~al.(2017)Dong, Huang, Wei, Lapata, Zhou, and Xu}]{ATT2Seq}
Dong, L.; Huang, S.; Wei, F.; Lapata, M.; Zhou, M.; and Xu, K. 2017.
\newblock Learning to Generate Product Reviews from Attributes.
\newblock In Lapata, M.; Blunsom, P.; and Koller, A., eds., \emph{ACL}, 623--632.

\bibitem[{Fu et~al.(2020)Fu, Xian, Gao, Zhao, Huang, Ge, Xu, Geng, Shah, Zhang, and de~Melo}]{KG}
Fu, Z.; Xian, Y.; Gao, R.; Zhao, J.; Huang, Q.; Ge, Y.; Xu, S.; Geng, S.; Shah, C.; Zhang, Y.; and de~Melo, G. 2020.
\newblock Fairness-Aware Explainable Recommendation over Knowledge Graphs.
\newblock In \emph{SIGIR}, 69–78. New York, NY, USA: Association for Computing Machinery.
\newblock ISBN 9781450380164.

\bibitem[{Hochreiter and Schmidhuber(1997)}]{lstm}
Hochreiter, S.; and Schmidhuber, J. 1997.
\newblock Long Short-Term Memory.
\newblock \emph{Neural Computation}, 9(8): 1735--1780.

\bibitem[{Hu et~al.(2022)Hu, Shen, Wallis, Allen-Zhu, Li, Wang, Wang, and Chen}]{lora}
Hu, E.~J.; Shen, Y.; Wallis, P.; Allen-Zhu, Z.; Li, Y.; Wang, S.; Wang, L.; and Chen, W. 2022.
\newblock Lo{RA}: Low-Rank Adaptation of Large Language Models.
\newblock In \emph{International Conference on Learning Representations}.

\bibitem[{Hu et~al.(2021)Hu, Ding, Wang, Liu, Wang, Li, Wu, and Sun}]{hu2021knowledgeable}
Hu, S.; Ding, N.; Wang, H.; Liu, Z.; Wang, J.; Li, J.; Wu, W.; and Sun, M. 2021.
\newblock Knowledgeable prompt-tuning: Incorporating knowledge into prompt verbalizer for text classification.
\newblock \emph{arXiv preprint arXiv:2108.02035}.

\bibitem[{Koren(2008)}]{SVD++}
Koren, Y. 2008.
\newblock Factorization meets the neighborhood: a multifaceted collaborative filtering model.
\newblock In \emph{SIGKDD}, 426–434. New York, NY, USA: Association for Computing Machinery.
\newblock ISBN 9781605581934.

\bibitem[{Li et~al.(2016)Li, Galley, Brockett, Gao, and Dolan}]{distinct-n}
Li, J.; Galley, M.; Brockett, C.; Gao, J.; and Dolan, B. 2016.
\newblock A Diversity-Promoting Objective Function for Neural Conversation Models.
\newblock In Knight, K.; Nenkova, A.; and Rambow, O., eds., \emph{Proceedings of the 2016 Conference of the North {A}merican Chapter of the Association for Computational Linguistics: Human Language Technologies}, 110--119. San Diego, California: Association for Computational Linguistics.

\bibitem[{Li et~al.(2020)Li, Chen, Zhang, Zhang, Zhang, Liu, and Ma}]{Sentires}
Li, L.; Chen, L.; Zhang, Y.; Zhang, H.; Zhang, M.; Liu, Y.; and Ma, S. 2020.
\newblock Sentires.
\newblock \url{https://github.com/lileipisces/Sentires-Guide}.

\bibitem[{Li, Zhang, and Chen(2020)}]{nete}
Li, L.; Zhang, Y.; and Chen, L. 2020.
\newblock Generate Neural Template Explanations for Recommendation.
\newblock In \emph{Proceedings of the 29th ACM International Conference on Information \& Knowledge Management}, CIKM '20, 755–764. New York, NY, USA: Association for Computing Machinery.
\newblock ISBN 9781450368599.

\bibitem[{Li, Zhang, and Chen(2021)}]{ACL21-PETER}
Li, L.; Zhang, Y.; and Chen, L. 2021.
\newblock Personalized Transformer for Explainable Recommendation.
\newblock In \emph{ACL}.

\bibitem[{Li, Zhang, and Chen(2023)}]{TOIS23-PEPLER}
Li, L.; Zhang, Y.; and Chen, L. 2023.
\newblock Personalized Prompt Learning for Explainable Recommendation.
\newblock \emph{ACM Transactions on Information Systems (TOIS)}.

\bibitem[{Li et~al.(2017)Li, Wang, Ren, Bing, and Lam}]{NRT}
Li, P.; Wang, Z.; Ren, Z.; Bing, L.; and Lam, W. 2017.
\newblock Neural Rating Regression with Abstractive Tips Generation for Recommendation.
\newblock In \emph{SIGIR}, 345–354.

\bibitem[{Lin(2004)}]{rouge}
Lin, C.-Y. 2004.
\newblock {ROUGE}: A Package for Automatic Evaluation of Summaries.
\newblock In \emph{Text Summarization Branches Out}, 74--81. Barcelona, Spain: Association for Computational Linguistics.

\bibitem[{Loshchilov and Hutter(2017)}]{Adamw}
Loshchilov, I.; and Hutter, F. 2017.
\newblock Decoupled Weight Decay Regularization.
\newblock In \emph{International Conference on Learning Representations}.

\bibitem[{Neelakantan et~al.(2022)Neelakantan, Xu, Puri, Radford, Han, Tworek, Yuan, Tezak, Kim, Hallacy et~al.}]{neelakantan2022text}
Neelakantan, A.; Xu, T.; Puri, R.; Radford, A.; Han, J.~M.; Tworek, J.; Yuan, Q.; Tezak, N.; Kim, J.~W.; Hallacy, C.; et~al. 2022.
\newblock Text and code embeddings by contrastive pre-training.
\newblock \emph{arXiv preprint arXiv:2201.10005}.

\bibitem[{Ni et~al.(2017)Ni, Lipton, Vikram, and McAuley}]{ni-etal-2017-estimating}
Ni, J.; Lipton, Z.~C.; Vikram, S.; and McAuley, J. 2017.
\newblock Estimating Reactions and Recommending Products with Generative Models of Reviews.
\newblock In Kondrak, G.; and Watanabe, T., eds., \emph{IJCNLP}, 783--791.

\bibitem[{{NLP Town}(2023)}]{sentiment}
{NLP Town}. 2023.
\newblock bert-base-multilingual-uncased-sentiment (Revision edd66ab).

\bibitem[{OpenAI(2023)}]{openai2023gpt4}
OpenAI. 2023.
\newblock GPT-4 Technical Report.
\newblock arXiv:2303.08774.

\bibitem[{Papineni et~al.(2002)Papineni, Roukos, Ward, and Zhu}]{bleu}
Papineni, K.; Roukos, S.; Ward, T.; and Zhu, W.-J. 2002.
\newblock {B}leu: a Method for Automatic Evaluation of Machine Translation.
\newblock In Isabelle, P.; Charniak, E.; and Lin, D., eds., \emph{Proceedings of the 40th Annual Meeting of the Association for Computational Linguistics}, 311--318. Philadelphia, Pennsylvania, USA: Association for Computational Linguistics.

\bibitem[{Raczyński, Lango, and Stefanowski(2023)}]{ECAI23-CER}
Raczyński, J.; Lango, M.; and Stefanowski, J. 2023.
\newblock The Problem of Coherence in Natural Language Explanations of Recommendations.
\newblock In \emph{ECAI}.

\bibitem[{Radford et~al.(2019)Radford, Wu, Child, Luan, Amodei, and Sutskever}]{Radford2019LanguageMA}
Radford, A.; Wu, J.; Child, R.; Luan, D.; Amodei, D.; and Sutskever, I. 2019.
\newblock Language Models are Unsupervised Multitask Learners.

\bibitem[{Sennrich, Haddow, and Birch(2016)}]{BPE}
Sennrich, R.; Haddow, B.; and Birch, A. 2016.
\newblock Neural Machine Translation of Rare Words with Subword Units.
\newblock In Erk, K.; and Smith, N.~A., eds., \emph{Proceedings of the 54th Annual Meeting of the Association for Computational Linguistics (Volume 1: Long Papers)}, 1715--1725. Berlin, Germany: Association for Computational Linguistics.

\bibitem[{Sun et~al.(2020)Sun, Wu, Zhang, Fu, Hong, and Wang}]{DualLearning}
Sun, P.; Wu, L.; Zhang, K.; Fu, Y.; Hong, R.; and Wang, M. 2020.
\newblock Dual Learning for Explainable Recommendation: Towards Unifying User Preference Prediction and Review Generation.
\newblock In \emph{WWW}, 837–847.

\bibitem[{Sun et~al.(2024)Sun, Shen, Zhou, Zhang, Chen, Cox, Yang, and Gan}]{10.5555/3666122.3666237}
Sun, Z.; Shen, Y.; Zhou, Q.; Zhang, H.; Chen, Z.; Cox, D.; Yang, Y.; and Gan, C. 2024.
\newblock Principle-driven self-alignment of language models from scratch with minimal human supervision.
\newblock In \emph{Proceedings of the 37th International Conference on Neural Information Processing Systems}, NIPS '23. Red Hook, NY, USA: Curran Associates Inc.

\bibitem[{Touvron et~al.(2023)Touvron, Lavril, Izacard, Martinet, Lachaux, Lacroix, Rozi{\`e}re, Goyal, Hambro, Azhar et~al.}]{touvron2023llama}
Touvron, H.; Lavril, T.; Izacard, G.; Martinet, X.; Lachaux, M.-A.; Lacroix, T.; Rozi{\`e}re, B.; Goyal, N.; Hambro, E.; Azhar, F.; et~al. 2023.
\newblock Llama: Open and efficient foundation language models.
\newblock \emph{arXiv preprint arXiv:2302.13971}.

\bibitem[{Vaswani et~al.(2017)Vaswani, Shazeer, Parmar, Uszkoreit, Jones, Gomez, Kaiser, and Polosukhin}]{transformer}
Vaswani, A.; Shazeer, N.; Parmar, N.; Uszkoreit, J.; Jones, L.; Gomez, A.~N.; Kaiser, L.; and Polosukhin, I. 2017.
\newblock Attention is all you need.
\newblock In \emph{Proceedings of the 31st International Conference on Neural Information Processing Systems}, NIPS'17, 6000–6010. Red Hook, NY, USA: Curran Associates Inc.
\newblock ISBN 9781510860964.

\bibitem[{Wang et~al.(2023)Wang, Yang, Huang, Yang, Majumder, and Wei}]{wang2023improving}
Wang, L.; Yang, N.; Huang, X.; Yang, L.; Majumder, R.; and Wei, F. 2023.
\newblock Improving text embeddings with large language models.
\newblock \emph{arXiv preprint arXiv:2401.00368}.

\bibitem[{Yang et~al.(2021)Yang, Wang, Deng, and Wang}]{yang2021explanation}
Yang, A.; Wang, N.; Deng, H.; and Wang, H. 2021.
\newblock Explanation as a Defense of Recommendation.
\newblock In \emph{Proceedings of the 14th ACM International Conference on Web Search and Data Mining}, 1029--1037.

\bibitem[{Zeng et~al.(2022)Zeng, Liu, Du, Wang, Lai, Ding, Yang, Xu, Zheng, Xia et~al.}]{zeng2022glm}
Zeng, A.; Liu, X.; Du, Z.; Wang, Z.; Lai, H.; Ding, M.; Yang, Z.; Xu, Y.; Zheng, W.; Xia, X.; et~al. 2022.
\newblock Glm-130b: An open bilingual pre-trained model.
\newblock \emph{arXiv preprint arXiv:2210.02414}.

\bibitem[{Zhang et~al.(2019)Zhang, Yao, Sun, and Tay}]{zhang2019deep}
Zhang, S.; Yao, L.; Sun, A.; and Tay, Y. 2019.
\newblock Deep learning based recommender system: A survey and new perspectives.
\newblock \emph{ACM Computing Surveys}, 52(1): 1--38.

\bibitem[{Zhang et~al.(2022)Zhang, Yan, Wang, and Wang}]{ZhangYWW22}
Zhang, W.; Yan, J.; Wang, Z.; and Wang, J. 2022.
\newblock Neuro-Symbolic Interpretable Collaborative Filtering for Attribute-based Recommendation.
\newblock In \emph{{WWW}}, 3229--3238.

\bibitem[{Zhang and Chen(2020)}]{ZhangC20}
Zhang, Y.; and Chen, X. 2020.
\newblock Explainable Recommendation: {A} Survey and New Perspectives.
\newblock \emph{Found. Trends Inf. Retr.}, 1--101.

\bibitem[{Zhang et~al.(2023)Zhang, Sun, Zhuang, Zhu, An, and Xu}]{zhang2023triple}
Zhang, Y.; Sun, Y.; Zhuang, F.; Zhu, Y.; An, Z.; and Xu, Y. 2023.
\newblock Triple Dual Learning for Opinion-based Explainable Recommendation.
\newblock \emph{ACM Transactions on Information Systems}, 42(3): 1--27.

\bibitem[{Zhang et~al.(2014)Zhang, Zhang, Zhang, Liu, and Ma}]{SIGIR14-Sentires}
Zhang, Y.; Zhang, H.; Zhang, M.; Liu, Y.; and Ma, S. 2014.
\newblock Do users rate or review? Boost phrase-level sentiment labeling with review-level sentiment classification.
\newblock In \emph{SIGIR}.

\bibitem[{Zheng, Noroozi, and Yu(2017)}]{DeepCoNN}
Zheng, L.; Noroozi, V.; and Yu, P.~S. 2017.
\newblock Joint deep modeling of users and items using reviews for recommendation.
\newblock In \emph{Proceedings of the tenth ACM international conference on web search and data mining}, 425--434.

\bibitem[{Zhou et~al.(2023)Zhou, Liu, Xu, Iyer, Sun, Mao, Ma, Efrat, Yu, YU, Zhang, Ghosh, Lewis, Zettlemoyer, and Levy}]{NEURIPS2023_ac662d74}
Zhou, C.; Liu, P.; Xu, P.; Iyer, S.; Sun, J.; Mao, Y.; Ma, X.; Efrat, A.; Yu, P.; YU, L.; Zhang, S.; Ghosh, G.; Lewis, M.; Zettlemoyer, L.; and Levy, O. 2023.
\newblock LIMA: Less Is More for Alignment.
\newblock In Oh, A.; Naumann, T.; Globerson, A.; Saenko, K.; Hardt, M.; and Levine, S., eds., \emph{Advances in Neural Information Processing Systems}, volume~36, 55006--55021. Curran Associates, Inc.

\end{thebibliography}

\appendix

\end{document}


\maketitle

\appendix

\section{Algorithm}
In order to better understand CIER, we provide the algorithm of CIER. Algorithm~\ref{alg:train} shows the process of training, and Algorithm~\ref{alg:infer} shows the process of inference.
\begin{algorithm}[!h]
    \caption{Algorithm of Training}
    \label{alg:train}
    \renewcommand{\algorithmicrequire}{\textbf{Input:}}
    \renewcommand{\algorithmicensure}{\textbf{Output:}}
    \begin{algorithmic}[1]
        \REQUIRE A training dataset $D$ includes user $u$, item $i$, corresponding rating $r$, explanation $e$ and backbone words $w$
        \ENSURE A trained CIER model    
        \FOR{each batch $(u, i, r, e, w) \in D$}
            \STATE Soft rating $r_s = Smooth(r)$ (Eq. (5))
            \STATE Rating embeddings $s_r = Embedding(r_s)$ (Eq. (3))
            \STATE Predicted rating distribution $\hat{r_s} = LLM(u,i)$ (Eq. (1))
            \STATE Compute $\mathcal{L}_r(r,\hat{r_s})$ (Eq. (8))
            \STATE Generate a random number $n \in [0,1]$
            \STATE Calculate $P(t)$ based on the current training step number $t$ (Eq. (6))
            \IF {$n \leq P(t)$}
                \STATE Generated explanation $\hat{e}=LLM(u, i, s_r)$ (Eq. (4))
                \STATE Compute $\mathcal{L}_e(e,\hat{e})$ (Eq. (9))
            \ELSE
                \STATE Generated backbone word $\hat{w}=LLM(u, i, s_r)$ (Eq. (4))
                \STATE Compute $\mathcal{L}_e(w,\hat{w})$ (Eq. (9))
            \ENDIF
            \STATE Update model by minimizing $\mathcal{L}_r$ and $\mathcal{L}_e$ (Eq. (10))
        \ENDFOR        
        
    \end{algorithmic}
\end{algorithm}

\begin{algorithm}[!h]
    \caption{Algorithm of Inference}
    \label{alg:infer}
    \renewcommand{\algorithmicrequire}{\textbf{Input:}}
    \renewcommand{\algorithmicensure}{\textbf{Output:}}
    \begin{algorithmic}[1]
        \REQUIRE A user $u$ and an item $i$
        \ENSURE A rating $\hat{r}$ and an explanation $\hat{e}$   
        
        \STATE Predicted rating distribution $\hat{r_s} = LLM(u,i)$ (Eq. (1))

        \STATE Calculate the rating $\hat{r} = \sum_{x=1}^|r| r_{s,i} \dot x$ (Eq. (2))

        \STATE Rating embeddings $s_r = E(\hat{r})$ (Eq. (3))
            
        \STATE Generated explanation $\hat{e}=LLM(u, i, s_r)$ (Eq. (4))  
        
    \end{algorithmic}
\end{algorithm}



\section{Comparing the results of using different PLMs as backbone.}
CIER does not strongly depend on a specific pre-trained language model. Table shows the effects of CIER on GPT-2, llama-v2-7b, and vicuna-7b-1.5. It can be found that on the larger dataset Yelp, the performance of gpt2 is not inferior to vicuna. On smaller datasets, gpt2 is difficult to match the 7b large-scale language model. LLM can better demonstrate its rich prior knowledge when the downstream dataset is smaller.
\begin{table}[h]
\small
    \begin{center}
        \resizebox{\linewidth}{!}{
            \begin{tabular}{l|cc|cc|cc}
                \hline
                \multirow{}{}&\multicolumn{2}{c|}{Yelp}&\multicolumn{2}{c|}{Amazon}&\multicolumn{2}{c}{TripAdvisor}\\
                \cline{2-7}
                &FMR$\uparrow$&R-L$\uparrow$&FMR$\uparrow$&R-L$\uparrow$&FMR$\uparrow$&R-L$\uparrow$\\
                \hline
                GPT-2&\textbf{8.87}&10.75&\textbf{12.49}&11.57&7,42&12.85\\
                Llama-v2-7b&8.71&\textbf{10.90}&12.45&11.70&\textbf{8.08}&13.40\\
                Vincuna-7b-1.5&8.70&10.76&12.42&\textbf{11.73}&7.97&\textbf{13.42}\\
                \hline
            \end{tabular}
        }
    \end{center}
    \caption{CIER's explainability and text quality on different PLMs.}
    \label{tab:plms}
\end{table}

\section{Comparing the results of different smoothing techniques}
The experimental results of applying different smoothing techniques on cier are shown in Table~\ref{tab:smooth}. When the rating prediction task is regarded as a classification task, the similarity between adjacent categories is high. Therefore, when smoothing, smoothing the rating to adjacent categories can bring better robustness to the model.
\begin{table}[h]
\small
    \begin{center}
        \resizebox{\linewidth}{!}{
            \begin{tabular}{l|cc|cc|cc}
                \hline
                \multirow{}{}&\multicolumn{2}{c|}{Yelp}&\multicolumn{2}{c|}{Amazon}&\multicolumn{2}{c}{TripAdvisor}\\
                \cline{2-7}
                &FMR$\uparrow$&R-L$\uparrow$&FMR$\uparrow$&R-L$\uparrow$&FMR$\uparrow$&R-L$\uparrow$\\
                \hline
                Hard rating&8.66&10.80&12.35&11.65&7.95&13.31\\
                Label Smoothing&8.63&10.87&12.33&11.61&8.01&13.36\\
                Gaussian smoothing&8.66&\textbf{10.91}&12.35&11.66&8.04&\textbf{13.40}\\
                Ours&\textbf{8.71}&10.90&\textbf{12.45}&\textbf{11.70}&\textbf{8.08}&\textbf{13.40}\\
                \hline
            \end{tabular}
        }
    \end{center}
    \caption{CIER's explainability and text quality on different smoothing techniques.}
    \label{tab:smooth}
\end{table}

\section{Instructions for using GPT-4 to automate evaluation of coherence}
Figure~\ref{fig:auto instruction} shows the instructions we designed for GPT-4 to enable it to automatically evaluate the coherence of predicted ratings and explanations. The instructions describe the rules for determining whether the ratings and explanations are coherent.
\begin{figure}[h]
    \centering
    \resizebox{\linewidth}{!}{
    \includegraphics{figures/auto instruction (2).pdf}
    }
    \caption{An instruction for GPT4 to perform automatic verification..}
    \label{fig:auto instruction}
\end{figure}